\documentclass[a4paper,prd,aps,floats,preprintnumbers,showpacs,twocolumn,amsmath,amssymb]{revtex4}
\pdfoutput=1
\usepackage{graphicx}
\usepackage{placeins}

\begin{document}

\preprint{
\vbox{
\hbox{ADP-09-12/T690}
}}

\title{Comparison of gluon flux-tube distributions for quark-diquark
and quark-antiquark hadrons } 

\author{F. Bissey}
\author{A. I. Signal}
\affiliation{Institute of Fundamental Sciences, Massey University,\\
        Private Bag 11 222, Palmerston North, New Zealand}
\author{D. B. Leinweber}
\affiliation{Centre for the Subatomic Structure of Matter and\\
        School of Chemistry and Physics, University of
Adelaide, SA 5005, Australia}


\begin{abstract}
The distribution of gluon fields in hadrons is of fundamental interest
in QCD.  
Using lattice QCD we have observed the formation of gluon flux tubes
within three quark 
(baryon) and quark plus antiquark (meson) systems for a wide variety
of spatial distributions 
of the color sources. 
In particular we have investigated three quark configurations where
two of the quarks are 
close together and the third quark is some distance away, which
approximates a quark 
plus diquark string.
We find that the string tension of the quark-diquark string is the
same as that of the 
quark-antiquark string on the same lattice.
We also compare the longitudinal and transverse profiles of the gluon
flux tubes for both 
sets of strings, and find them to be of similar radii and to have
similar vacuum suppression.

\end{abstract}


\maketitle

\section{INTRODUCTION}

There has recently been a renewal of interest in the properties of
diquarks in hadronic systems, as
they may play an important role in the existence of exotic states,
such as the $\Theta^{+}$, or in 
explaining the scarcity of such exotics \cite{JaffeWilczek}. 
While string-type models consisting of a quark plus a diquark have
been studied using various 
analytic techniques (see \cite{Anselmino93} for a review), there have
only been a few studies of 
this type of system on the lattice. 
These studies have investigated the mass of diquarks 
\cite{Hess98,Babich05,Orginos05,AdeFL06,Liu06}, 
and, more recently, the nature of diquark correlations
\cite{Babich07}.
Recently we investigated the formation of flux tubes in static baryon
systems on the lattice 
using a high statistics approach which enabled us to observe
correlations between the 
vacuum action density and the quark positions in a gauge
independent manner 
\cite{Bissey:2005sk} 
In that work the three quarks were positioned approximately
equidistant, and a Y-shaped 
flux-tube was observed to form at large inter-quark distances. 
In this work we extend our study of three quark systems to the case
where two of the quarks are 
close together and the third is some distance away. 

In QCD, two quarks close together, a diquark, can transform either
according to the conjugate 
representation $(\bar{3})$ or the sextet $(6)$ representation of
$SU(3)$. 
The color hyperfine interaction then leads to attraction in the spin
singlet, scalar diquark 
channel, while the spin triplet, axial vector diquark is disfavoured. 
Hence low-lying diquarks should have positive parity and belong to the
color $\bar{3}$ representation, and 
so have many properties similar to an antiquark. 
In lattice QCD this should lead to the formation of quark-diquark
flux tubes with similar physical 
characteristics to those of quark-antiquark flux tubes. 
In particular we would expect the long range linear part of the 
quark-diquark potential to have the 
same slope as that of the quark-antiquark potential, corresponding
to the flux tubes having the 
same energy density, and we would expect the flux tubes to have
similar transverse size. 
In this work we investigate whether these similarities do indeed hold.

\section{Flux tubes on the Lattice}

\begin{figure}
\centering\includegraphics[height=10cm,clip=true]{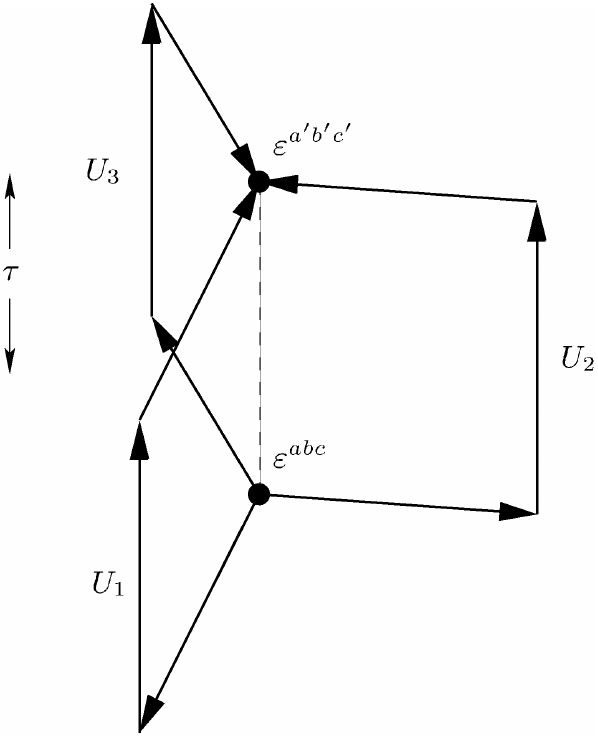}
\caption{Gauge-link paths or ``staples,'' $U_1$, $U_2$ and $U_3$,
  forming a three-quark Wilson loop with the quarks located at $\vec
  r_1$, $\vec r_2$ and $\vec r_3$.  $\varepsilon^{abc}$ and
  $\varepsilon^{a'b'c'}$ denote colour anti-symmetrisation at the
source
  and sink respectively, while $\tau$ indicates evolution of the
  three-quark system in Euclidean time.}
\label{staple}
\end{figure}

In order to study flux-tubes on the lattice, we begin with the
standard approach of connecting 
static quark (and antiquark) propagators with spatial-link paths in a
gauge invariant manner.  
We use APE-smeared spatial-link paths to propagate the quarks from a
common origin to their
spatial positions as illustrated in Fig.~\ref{staple}.  
In earlier work we saw that after approximately 30 APE smearing steps we had
obtained optimal overlap
with the ground state. 
The static quark propagators are constructed from time directed link
products at fixed spatial 
coordinate, $\prod_i U_t(\vec x, t_i)$, using the untouched `thin'
links of the gauge configuration.
In principle, the ground state is isolated after sufficient time
evolution.  
Finally smeared-link spatial paths propagate the quarks back to the
common
spatial origin.

\begin{figure}
\centering\includegraphics[width=8cm,clip=true]{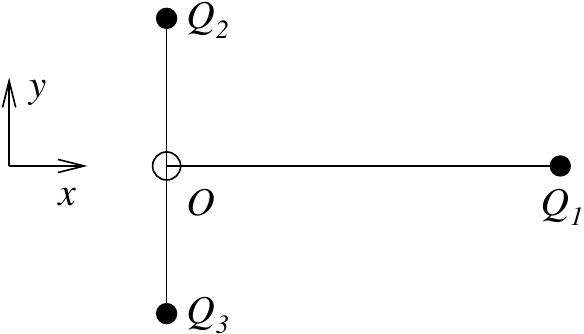}
\caption{Projection of the T-shape path on the $x$-$y$ plane.}
\label{tpath}
\end{figure}

The three-quark Wilson loop is thus defined as:
\begin{equation}
W_{3Q}=\frac{1}{3!}\varepsilon^{abc}\varepsilon^{a'b'c'} \, U_1^{aa'}
\,
U_2^{bb'} \, U_3^{cc'},
\end{equation}  
where $U_j$ is a staple made of path-ordered link variables
\begin{equation}
U_j \equiv P \exp \left( ig \int_{\Gamma_j} dx_{\mu} \, A^{\mu}(x)
\right) \, ,
\end{equation}
and $\Gamma_j$ is the path along a given staple as shown in 
Fig.~\ref{staple}.
In contrast the quark-antiquark Wilson loop is given by the product
of two staples
\begin{equation}
W_{Q\bar{Q}} = \delta^{ab} \delta^{a'b'} \, U_{1}^{aa'} \,
\left(U_{2}^{bb'}\right)^{\dagger}.
\end{equation}
The three-quark configurations we use to approximate a quark-diquark
string are T-shapes, 
with the origin at the junction of the T. Two quarks are positioned
one lattice step in opposite 
directions from the origin (approximating the diquark), and the third
is placed from 1 to 12
lattice steps in an orthogonal direction, as shown in
Fig.~\ref{tpath}.

In this work we have used 300 quenched QCD gauge field configurations
created with the 
${\cal O}(a^2)$-mean-field improved Luscher-Weisz plaquette plus
rectangle gauge action
\cite{Luscher:1984xn} on $16^3\times 32$ lattices with the long
dimension being the 
$x$ direction, making the spatial volume $16^2\times 32$.
Two hundred of these configurations  were at at $\beta =4.60$ (as in
our previous work) and 
one hundred at $\beta = 4.80$, to investigate the use of a finer
lattice.
These values of $\beta$ give lattice spacings $a$ of $0.123$ fm and
$0.0945$ fm respectively.

We use lattice symmetries to improve the signal to noise ratio of our
measurements. 
These include translational invariance (any point on the lattice can
be taken as the origin), 
reflection in the $x$ plane and $90^\circ$ rotational symmetry about
the $x$-axis.
The advantage of this approach is that we do not have to perform any
gauge fixing to find a 
signal in our flux distributions. 

We characterise the gluon-field fluctuations in our configurations
using the gauge-invariant 
action density $S(\vec y, t)$ observed at spatial coordinate $\vec y$
and Euclidean time $t$ 
relative to the origin of the Wilson loop.  
We calculate the action density using the highly-improved ${\cal
O}(a^4)$ three-loop improved
lattice field-strength tensor \cite{Bilson-Thompson:2002jk} on
four-sweep APE-smeared gauge 
links.  
While the use of this highly-improved action suppresses correlations
close to the quark 
positions, it gives good resolution of the flux-tube correlations we
are interested in. 

Defining the quark positions as $\vec r_i$ relative to the origin of
the Wilson loop,
and denoting the Euclidean time extent of the loop by $\tau$, we
evaluate
the following correlation functions
\begin{eqnarray}
C_{3Q}(\vec y; \vec r_1, \vec r_2, \vec r_3; \tau)\!\! & = &\!\!
\frac{
\bigl\langle W_{3Q}(\vec r_1, \vec r_2, \vec r_3; \tau) \,
             S(\vec y, \tau/2) \bigr\rangle }
{
\bigl\langle  W_{3Q}(
\tau) \bigr\rangle \,
\bigl\langle S(\vec y, \tau/2) \bigr\rangle
},  \\
C_{Q\bar{Q}}(\vec y; \vec r_1, \vec r_2; \tau) & = &
 \frac{
\bigl\langle W_{Q\bar{Q}}(\vec r_1, \vec r_2; \tau) \,
             S(\vec y, \tau/2) \bigr\rangle }
{
\bigl\langle  W_{Q\bar{Q}}(
\tau) \bigr\rangle \,
\bigl\langle S(\vec y, \tau/2) \bigr\rangle
},
\label{correl}
\end{eqnarray} 
where $\langle \cdots \rangle$ denotes averaging over configurations
and lattice symmetries.
These correlate the quark positions, via the Wilson loops, with the
gauge-field action in a 
gauge invariant manner.  
For fixed quark positions and Euclidean time, the correlation
functions are scalar fields in
three dimensions.  
For values of $\vec y$ well away from the quark positions $\vec r_i$,
there are no correlations 
and $C \to 1$.
Also the correlators are positive definite, eliminating any sign
ambiguity on whether vacuum 
field fluctuations are enhanced or suppressed in the presence of
static quarks.  
We find that $C$ is generally less than 1, signaling the expulsion of
vacuum fluctuations
from the interior of hadrons.

\begin{figure}
\centering\includegraphics[height=6cm,clip=true]{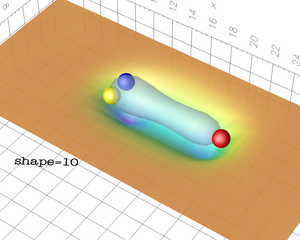}
\caption{Expulsion of gluon-field fluctuations from the region of
  static quark sources illustrated by the spheres.  An isosurface of
  $C(\vec{y})$ is illustrated by the translucent surface.  A
  surface plot (or rubber sheet) describes the values of $C(\vec{y})$
  for $\vec y$ in the quark plane, $(y_1, y_2, 0)$.}
\label{T-tube10}
\end{figure}

\begin{figure}
\centering\includegraphics[height=6cm,clip=true]{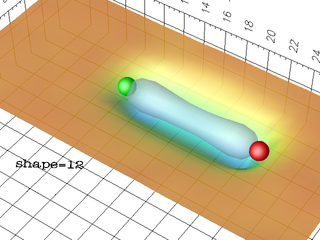}
\caption{Formation of quark-antiquark flux-tube. Details are
as in \ref{T-tube10}}
\label{I-tube12}
\end{figure}

In Figs.~\ref{T-tube10} and ~\ref{I-tube12} we show examples of the
expulsion of vacuum 
fluctuations and the formation of flux-tubes for our quark-diquark
and quark-antiquark 
configurations.

\section{Effective Potentials}

\begin{figure}
\centering\includegraphics[height=6cm,clip=true]{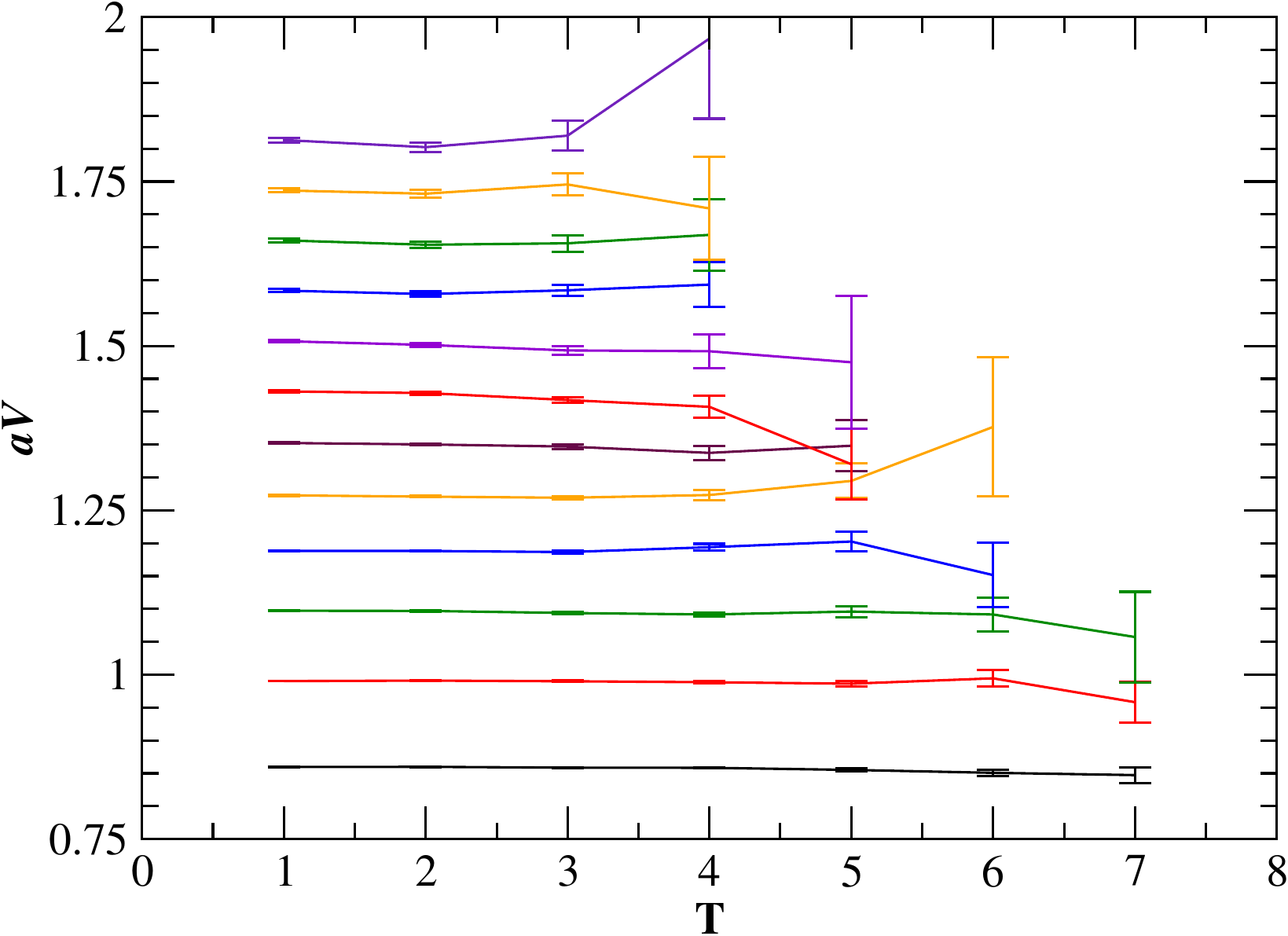}
\centering\includegraphics[height=6cm,clip=true]{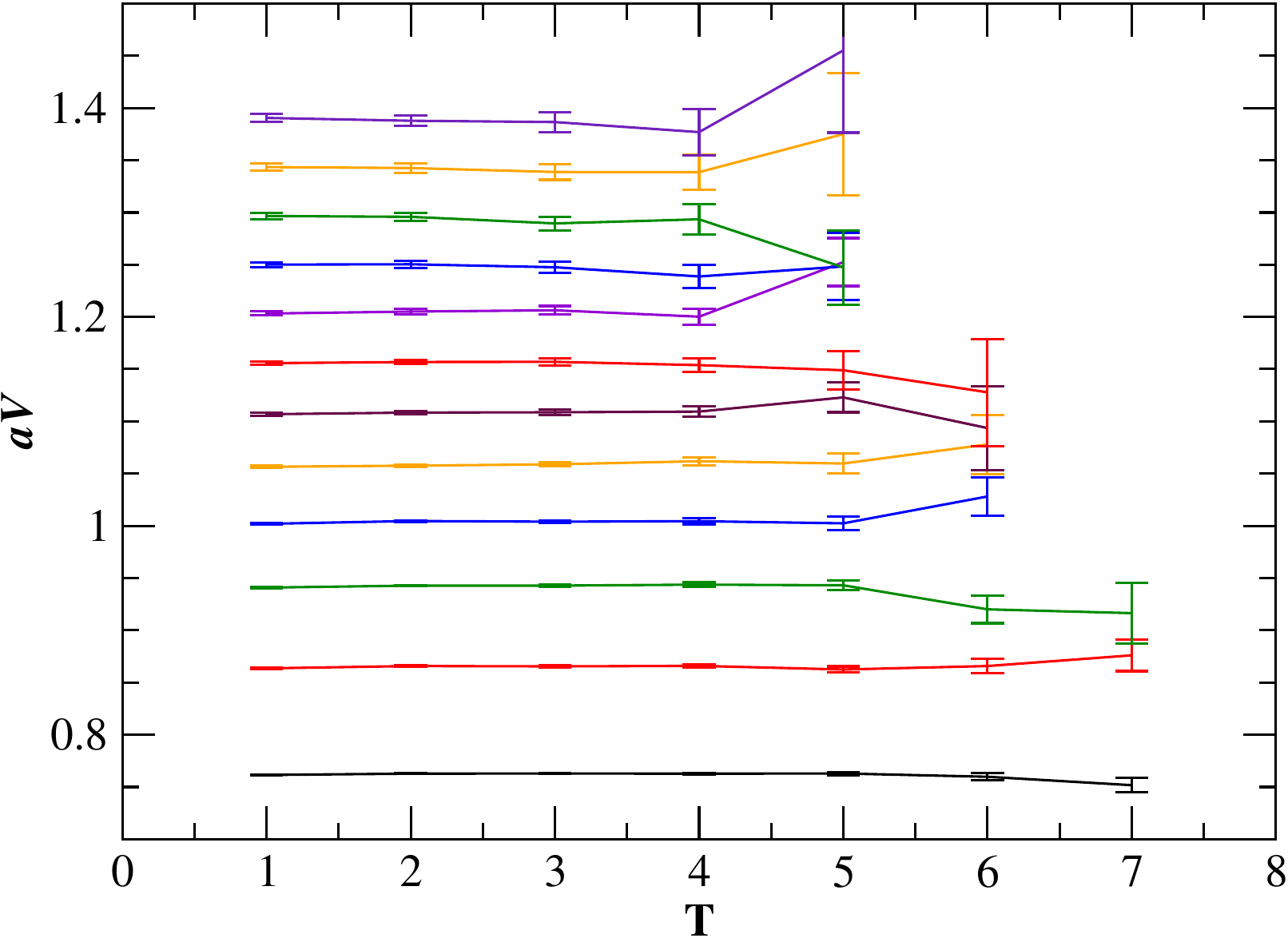}
\caption{Effective static quark potential for 30-sweep smeared 
quark-diquark sources for 
$\beta = 4.6$ (top) and $\beta = 4.8$ (bottom).  
From bottom up, the lines correspond to quark-diquark separation
increasing from 1 to 12 lattice spacings.}
\label{plat}
\end{figure}

\begin{figure}
\centering\includegraphics[height=6cm,clip=true]{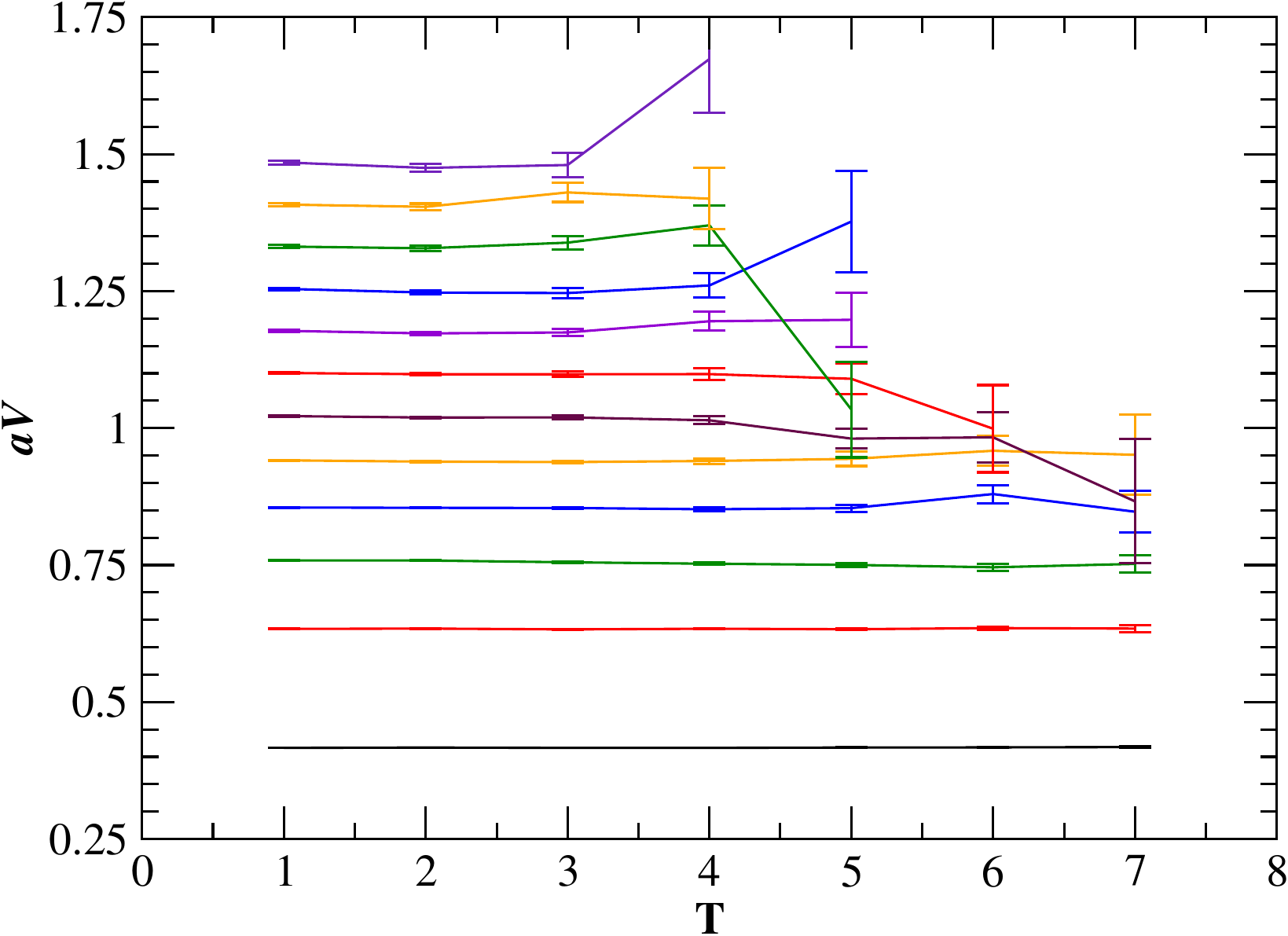}
\centering\includegraphics[height=6cm,clip=true]{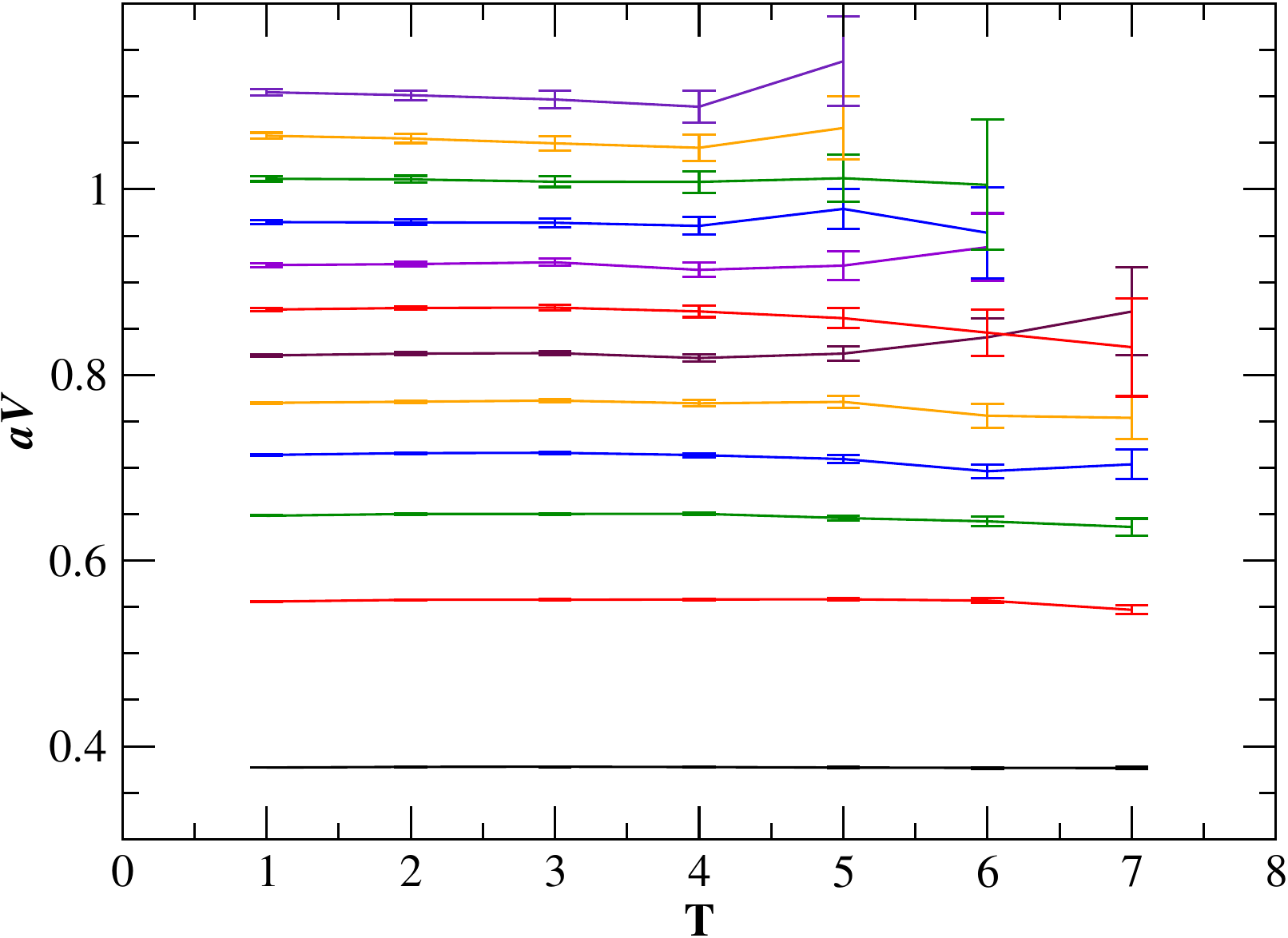}
\caption{Effective static quark potential for 30-sweep smeared 
quark-antiquark sources for 
$\beta = 4.6$ (top) and $\beta = 4.8$ (bottom).  
From bottom up, the lines correspond to quark-diquark separation
increasing from 1 to 12 lattice spacings.}
\label{plat2}
\end{figure}

In this section we extract the long range portion of the potential for
our quark-antiquark and 
quark-diquark  flux-tubes. 
According to QCD, the effective potentials should have the same slope,
so long as the APE 
smearing in the spatial directions has smoothed the gluon fluctuations
sufficiently to isolate 
the ground state and the propagation of the Wilson loops in the time
direction is long enough 
for any excited states to decay.
The effective potential is obtained from the Wilson loops in the
standard manner: 
\begin{equation}
a\, V(\vec{r},\tau)= \ln\left(
\frac{W(\vec{r},\tau)}{W(\vec{r},\tau+1)}\right).
\label{pot}
\end{equation}
As shown in Fig.~\ref{plat} and \ref{plat2}, we obtain stable plateaus
for the potentials as a 
function of $\tau$. 
The statistical uncertainties are estimated using the jackknife method
\cite{Montvay:1994bk}. 


\begin{figure}
\centering\includegraphics[width=8cm,clip=true]{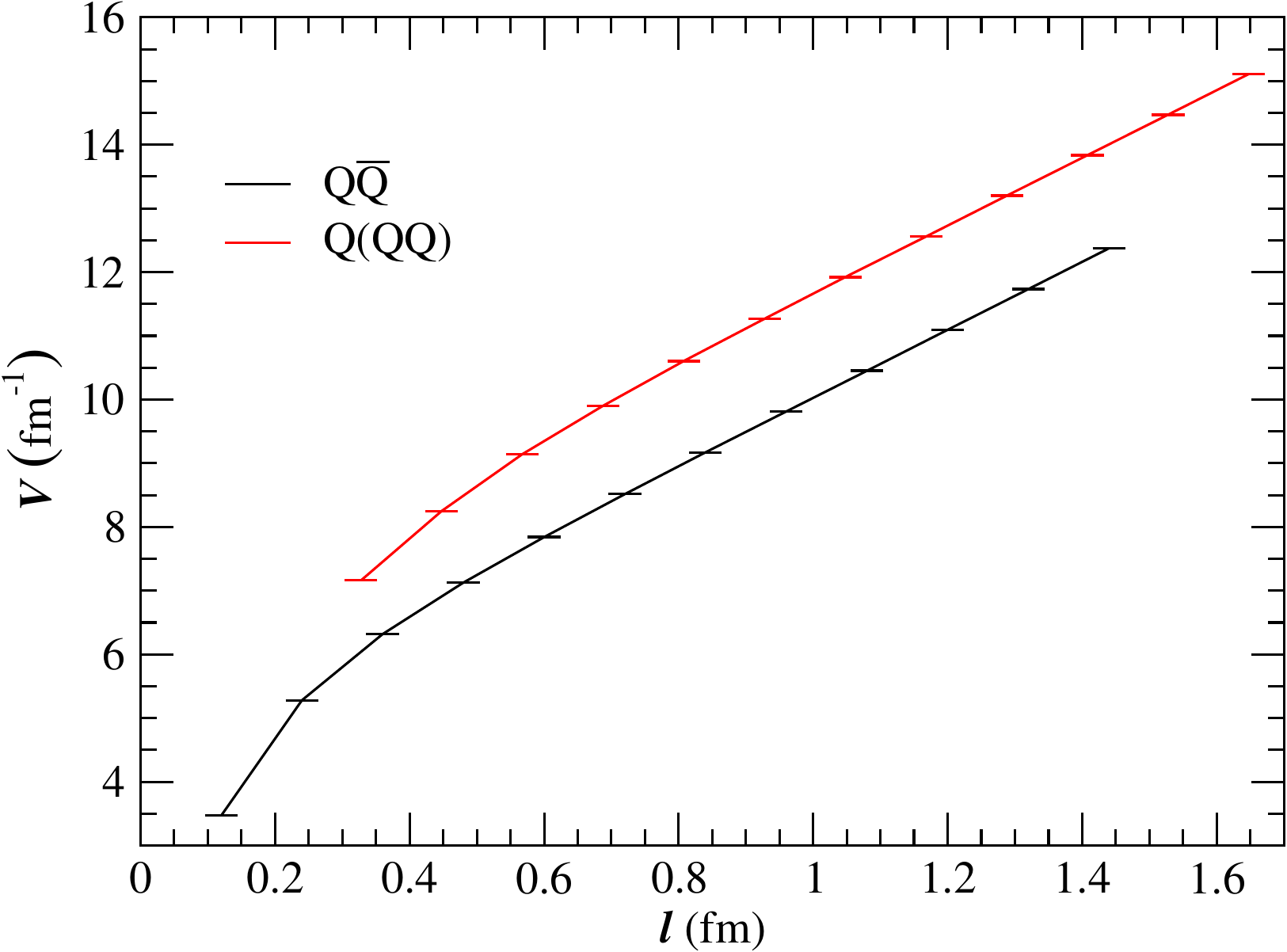}
\centering\includegraphics[width=8cm,clip=true]{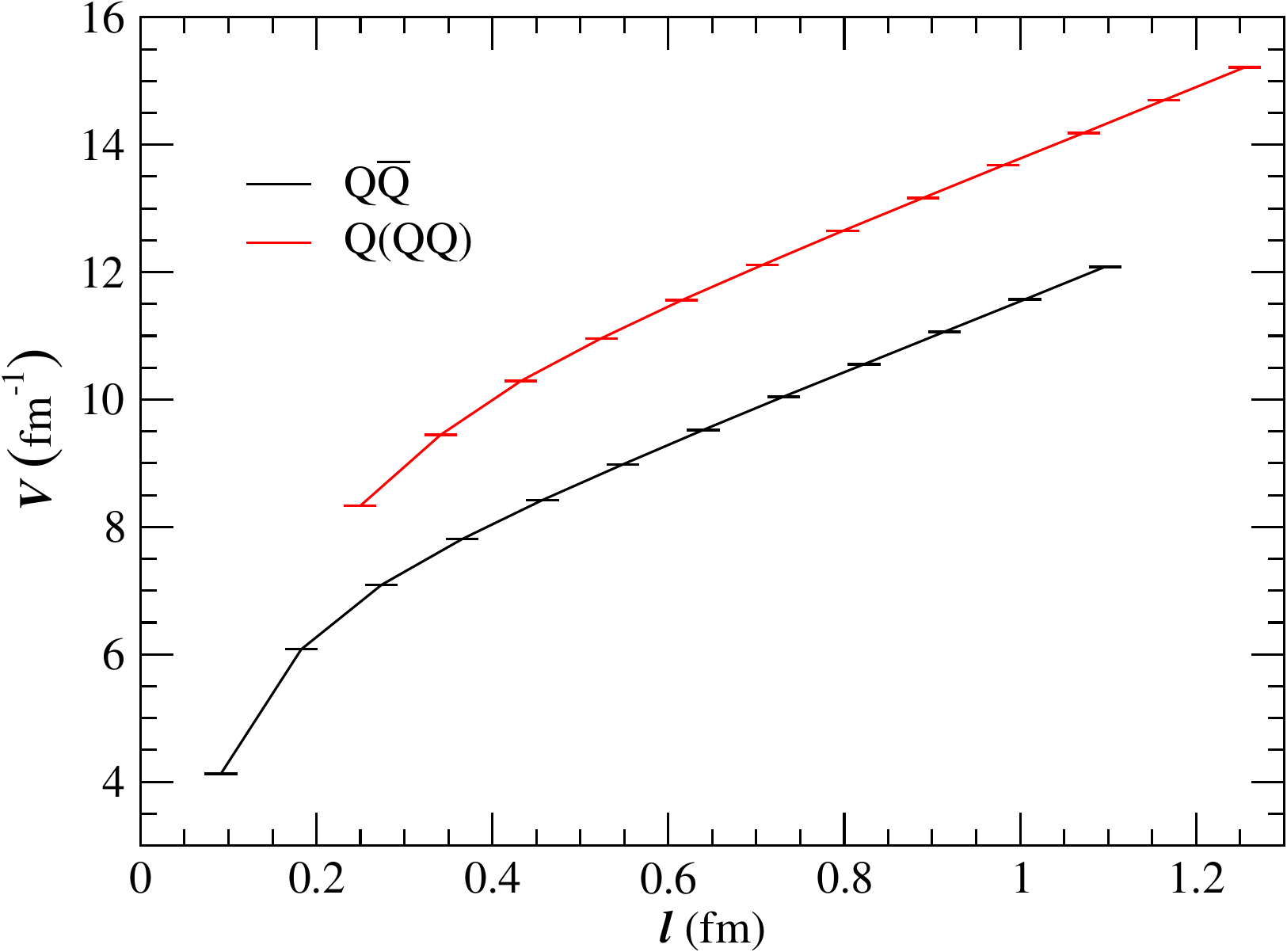}
\caption{Comparison of quark-antiquark and quark-diquark effective
potentials for 
$\beta = 4.6$ (top) and $\beta = 4.8$ (bottom).}
\label{effpots}
\end{figure}

The quark-antiquark potential  has the well-known form
\begin{equation}
V_{Q\bar{Q}}(r) = V_{0} - \frac{\alpha}{r} + \sigma_{Q\bar{Q}} r,
\label{potform2}
\end{equation}
where $\sigma$ is the string tension.
The three quark potential is \cite{Takahashi:2002bw,Alexandrou:2001ip}
\begin{equation}
V_{3Q}(r) = \frac{3}{2}V_0 -\frac{1}{2}\sum_{j<k} \frac{g^2 C_F}{4\pi
r_{jk}}
+\sigma_{3Q} L(r) \, ,
\label{potform3}
\end{equation}
where $C_F =4/3$ and $L(r)$ is a length linking the quarks.  
As shown in our earlier work \cite{Bissey:2005sk}, $L(r)$ is given by
the minimum length of string 
that connects the three quarks, or the sum of distances from the
quarks to the Fermat (or Steiner) 
point.
QCD suggests that the two string tensions $\sigma_{Q\bar{Q}}$ and
$\sigma_{3Q}$ are equal.
In Fig.~\ref{effpots} we plot the extracted effective potentials for
the quark-diquark and 
quark-antiquark flux-tubes at each of the values of $\beta$ for our
gauge configurations. 
The plots in fermi show that the QCD prediction is confirmed
at both values of $\beta$. 
Converting length measurements from lattice units to fermi we obtain
the quark-diquark string 
tension $\sigma_{3Q} = 0.97 \pm 0.01$ GeV fm$^{-1}$, which is in
excellent agreement with the quark-antiquark string tension
$\sigma_{Q\bar{Q}} = 0.98$ GeV fm$^{-1}$, defining the lattice spacing.

\section{Flux-tube Profiles}

We can gain further insight into the properties of the flux tubes by
examining 
their profiles close to the quark.
We study the values of the correlators $C_{3Q}(\vec{y})$ and
$C_{Q\bar{Q}}(\vec{y})$ where
$\vec{y} = (y_{1}, y_{2}, 0)$ is constrained to the plane of the
color sources, and the origin is at 
the position of either the antiquark or the join of the T. 
The quark is then at the position $(\xi, 0, 0)$ where $\xi$ varies
from 1 to 12 lattice steps.

\begin{figure}
\centering\includegraphics[width=8cm,clip=true]{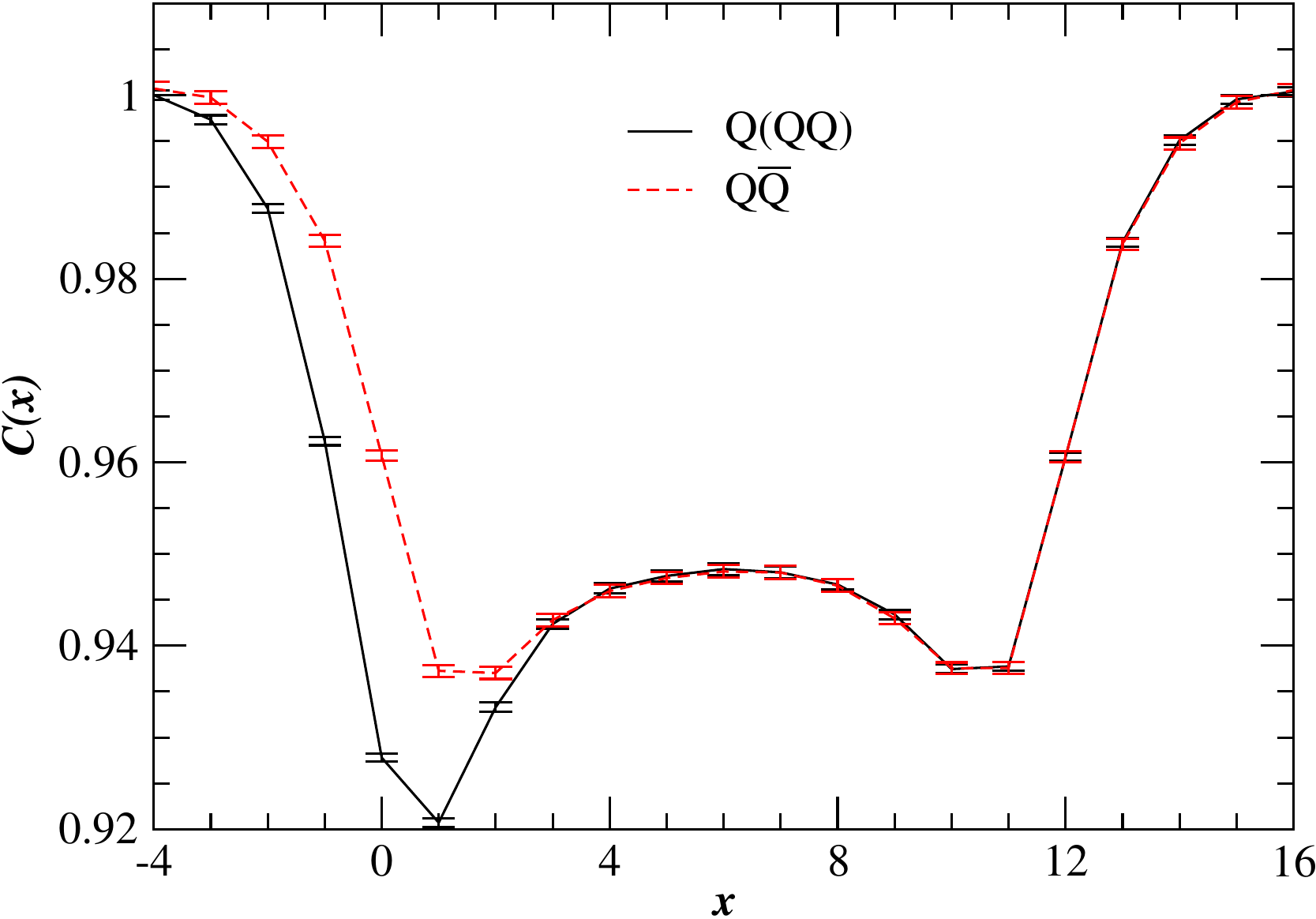}
\caption{Comparison of longitudinal flux tube profiles for $\beta =
4.8$ at longitudinal 
separation of 12 lattice units.}
\label{longprofiles}
\end{figure}

First we examine the longitudinal profiles of both quark-diquark and
quark-antiquark 
flux-tubes along the line $(\vec{y}) = (x, 0, 0)$ in
Fig.~\ref{longprofiles}. 
As expected, the vacuum expulsion close to the diquark is stronger
than in the vicinity of the 
antiquark. 
However, near the quark the two flux tubes show very similar
profiles. 
Similar results are seen at $\beta = 4.6$.

\begin{figure}
\centering\includegraphics[width=8cm,clip=true]{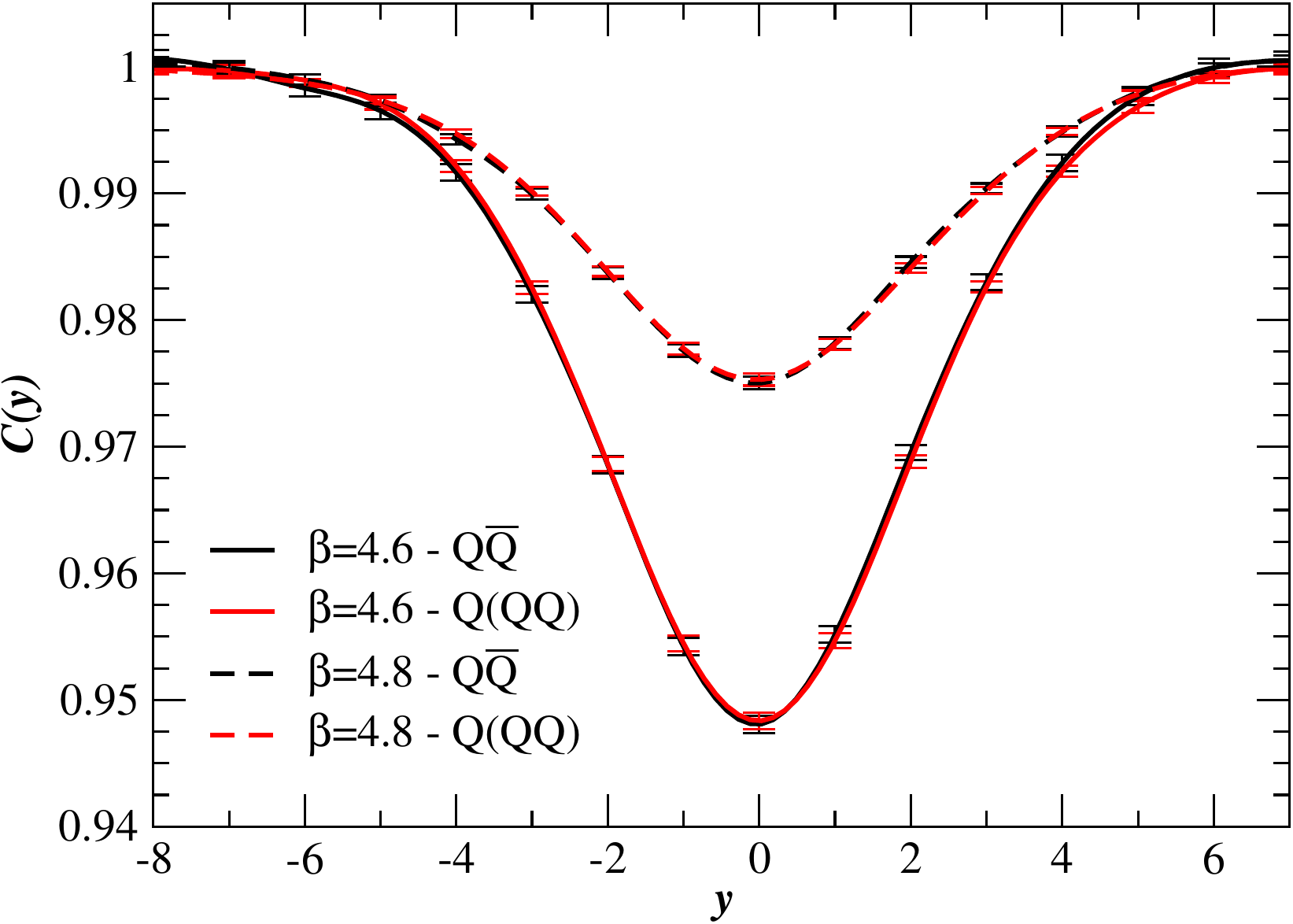}
\caption{Transverse profiles for quark-diquark (red lines) and 
quark-antiquark (black lines) 
flux tubes at $\beta = 4.6$ (solid lines) and $\beta = 4.8$ (dashed
lines).}
\label{trprofiles}
\end{figure}

Next we examine the transverse profiles along a line orthogonal to the
midpoint of the flux tube, 
{\em ie.} along $(\xi /2, y, 0)$ for $\xi$ even, or along  $((\xi +1)
/2, y, 0)$ for $\xi$ odd.
In Fig.~\ref{trprofiles} we show profiles of both quark-diquark and
quark-antiquark flux-tubes 
for $\xi = 12$. 
We find that as long as $\xi$ is larger than one third 
of the total length of the quark-antiquark system, the
transverse profiles are close to identical.

The transverse profile of the flux-tube is fitted well by a Gaussian
function
$C(y) = 1 - A e^{-y^{2}/r^{2}}$. 
The fit enables us to estimate the radius $(r)$ and area
$(Ar\sqrt{\pi})$ of the flux-tube as well 
as its depth. 
The fit parameters for our flux tubes are given in
Table~\ref{flux-fit}. 
Again, the fits show that for long enough flux tubes the transverse
profiles of quark-diquark and 
quark-antiquark flux-tubes are statistically identical.

\begin{table}
\caption{Values of the fit parameters of the function $1-A\,
  \exp(-y^2/r^2)$ to the transverse profiles of the flux tubes. 
  $r$ is reported in lattice units (LU) and
  fm.  The last column indicates the area under $C(\vec{y})=1$ for the
  fitted curve in units of relative-depth times LU.}
\label{flux-fit}
\begin{ruledtabular}
\begin{tabular}{cccccc}
Flux Tube     & $\beta$   & $A$        & $r$ (LU)   & $r$ (fm)  & $Ar
\sqrt{\pi}$ \\
\noalign{\smallskip}
\hline
\noalign{\smallskip}
$Q \bar{Q}$ & $4.6$ & $0.0511(1)$ & $2.89(3)$   & $0.355(3)$ &
$0.261(3)$ \\
$QQQ$         & $4.6$ & $0.0510(1)$ & $2.91(2)$   & $0.358(3)$ &
$0.263(3)$ \\
\hline
$Q \bar{Q}$ & $4.8$ & $ 0.0243(3) $ & $3.17(4)$   & $0.299(4)$ &
$0.136(3)$ \\
$QQQ$         & $4.8$ & $0.0241(3)$ & $3.21(3)$   & $0.303(3)$ &
$0.137(3)$ \\
\end{tabular}
\end{ruledtabular}
\end{table}

\section{Conclusions}

We have directly compared gluon flux-tubes for quark plus antiquark
and three quark systems. 
In the three quark systems we kept two quarks close together (two
lattice units separation),
so that the system would approximate a quark-diquark string. 
We found that the string tension in the quark-diquark string was the
same as for the 
quark-antiquark string.
In addition we compared the vacuum expulsion in both sets of
flux-tubes. 
We found that, in the vicinity of the quark, there was no measurable
difference between the  
transverse profiles of the quark-diquark flux-tubes and the 
quark-antiquark flux-tubes. 
Also the longitudinal profiles of both sets of flux-tubes were very
similar. 

These findings confirm the expectation from QCD that a diquark has
many properties in common 
with an antiquark.
In particular the long range color interaction between a diquark and a
quark is seen to be the 
same as that between an antiquark and a quark.
This result is interesting in that it is obtained in the quenched
approximation, where the color 
hyperfine interaction should be small. 
This implies that the APE smearing and propagation in Euclidean time
we have performed has 
been sufficient for decuplet baryon states to decay. 
It would be interesting to repeat this work with dynamical quarks,
where variation in the strength 
of the color hyperfine interaction could be investigated.
This is potentially of great importance to phenomenological models of
hadron structure.

\bigskip

\begin{acknowledgments}
This work has been done using the DoubleHelix computer at 
Massey University and supercomputing resources from the 
NCI National Facility and eResearch SA. The 3-D
realisations have been rendered using OpenDX
(http://www.opendx.org). The 2D plots and curve fitting have been done
using Grace (http://plasma-gate.weizmann.ac.il/Grace/).
\end{acknowledgments}


\begin{thebibliography}{9}

\bibitem{JaffeWilczek}
R.~L.~Jaffe and F.~Wilczek,
Phys. Rev. Lett. {\bf 91}, 232003 (2003)
[arXiv:hep-ph/0307341].
R.~L.~Jaffe,
Phys. Rept. {\bf 409}, 1 (2005)
[arXiv:hep-ph/0409065].

\bibitem{Anselmino93}
M.~Anselmino, E~. Predazzi, S.~Ekelin, S.~Fredriksson and
D.~B.~Lictenberg,
Rev. Mod. Phys. {\bf 65}, 1199 (1993).

\bibitem{Hess98}
M.~Hess, F.~Karsch, E.~Laermann, and I.~Wetzorke 
Phys. Rev. D {\bf 58}, 111502 (1998)
[arXiv:hep-lat/9804023].
 
 \bibitem{Babich05}
quarks on a large lattice.
L.~Lellouch, C.~Rebbi and N.~Shoresh 
 R.~Babich {\it et al.}
JHEP {\bf 0601} 086 (2006)
[arXiv:hep-lat/0509027].
  
\bibitem{Orginos05}
K.~Orginos 
Proc. Science LAT2005, 054 (2006)
[arXiv:hep-lat/0510082].
   
\bibitem{AdeFL06}
C.~Alexandrou, Ph.~de Forcrand and B.~Lucini 
Phys. Rev. Lett. {\bf97}, 222002 (2006)
[arXiv:hep-lat/0609004].
 
\bibitem{Liu06} 
simulations.
Z.~Liu and T.~DeGrand 
Proc. Science LAT2006 116 (2006)
[arXiv:hep-lat/0609038].

\bibitem{Babich07}
R.~Babich {\it et al.}
Christian Hoelbling , 
[arXiv:hep-lat/0701023].



\bibitem{Takahashi:2002bw}
  T.~T.~Takahashi, H.~Matsufuru, Y.~Nemoto and H.~Suganuma,
  Phys. Rev. Lett. \textbf{86}, 18 (2001).
  T.~T.~Takahashi, H.~Suganuma, Y.~Nemoto and H.~Matsufuru,
QCD,''
  Phys.\ Rev.\ D {\bf 65}, 114509 (2002)
  [arXiv:hep-lat/0204011].

\bibitem{Alexandrou:2001ip}
  C.~Alexandrou, P.~De Forcrand and A.~Tsapalis,
  Phys.\ Rev.\ D {\bf 65}, 054503 (2002)
  [arXiv:hep-lat/0107006].
  C.~Alexandrou, P.~de Forcrand and O.~Jahn,
  Nucl.\ Phys.\ Proc.\ Suppl.\  {\bf 119}, 667 (2003)
  [arXiv:hep-lat/0209062].





\bibitem{Bissey:2005sk}
  F.~Bissey {\it et al.}
  Nucl.\ Phys.\ Proc.\ Suppl.\  {\bf 141}, 22 (2005)
  [arXiv:hep-lat/0501004];
  F.~Bissey {\it et al.},
baryons,''
  Phys. Rev. D {\bf 76}, 114512 (2007) 
  [arXiv:hep-lat/0606016].





\bibitem{Bilson-Thompson:2002jk}
S.~O.~Bilson-Thompson, D.~B.~Leinweber and A.~G.~Williams,
Annals Phys.\  {\bf 304}, 1 (2003)
[hep-lat/0203008].

\bibitem{Luscher:1984xn}
M.~Luscher and P.~Weisz,
Commun.\ Math.\ Phys.\  {\bf 97}, 59 (1985)
[Erratum-ibid.\  {\bf 98}, 433 (1985)].

\bibitem{Montvay:1994bk}
I. Montvay and G. M\"unster, ``Quantum Fields on a Lattice''
(Cambridge, 1994) 389.



\end{thebibliography}
\end{document}